\documentstyle{article}
\begin{document}
\font\cal=cmsy10 at 10 pt

\def\a{\alpha}
\def\A{\Alpha}
\def\Alpha{\Arm}
\def\b{\beta}
\def\B{{\Beta}}
\def\Chi{\Xrm}
\def\D{\Delta}
\def\d{\delta}
\def\e{\epsilon}
\def\F{\Phi}
\def\f{\phi}
\def\G{\Gamma}
\def\g{\gamma}
\def\i{\iota}
\def\vf{\varphi}
\def\k{\kappa}
\def\L{\Lambda}
\def\l{\lambda}
\def\M{\Mu}
\def\Mu{\Mrm}
\def\m{\mu}
\def\n{\nu}
\def\nab{\nabla}
\def\N{\Nu}
\def\Nu{\Nrm}
\def\p{\pi}
\def\vr{\varrho}
\def\r{\rho}
\def\Si{\Sigma}
\def\si{\sigma}
\def\s{\sigma}
\def\t{\tau}
\def\Th{\Theta}
\def\th{\theta}
\def\W{\Om}
\def\w{\omega}
\def\Y{\Psi}
\def\y{\psi}
\def\z{\zeta}


\def\Csfs{{\hbox{\sfs C}}} 
\def\Csf{\hbox{\sf C}}
\def\Dsf{\hbox{\sf D}}
\def\Dsfs{\hbox{\sfs D}}
\def\Esf{\hbox{\sf E}}
\def\Fsfs{\hbox{\sfs F}}
\def\Fsf{\hbox{\sf F}}
\def\Gsfs{\hbox{\sfs G}}
\def\Gsf{\hbox{\sf G}}
\def\gsf{\hbox{\sf g}}
\def\gsfs{\hbox{\sfs g}}
\def\hsf{\hbox{\sf h}}
\def\Hsfs{{\hbox{\sfs H}}}
\def\Hsf{\hbox{\sf H}}
\def\Isf{\hbox{\sf I}}
\def\Isfs{\hbox{\sfs I}}
\def\Msfs{{\hbox{\sfs M}}}
\def\Msf{\hbox{\sf M}}
\def\Nsf{\hbox{\sf N}}
\def\Rsf{\hbox{\sf R}}
\def\tsf{\hbox{\sf t}}
\def\Tsfs{{\hbox{\sfs T}}}
\def\Tsf{\hbox{\sf T}}
\def\Tsfp{\hbox{\sfp T}}

\def\Vsf{\hbox{\sf V}}
\def\Xsf{\hbox{\sf X}}
\def\Zsf{\hbox{\sf Z}}

\def\Qb{{\bf Q}}


\def\Cbb{\Isf{}\kern-4pt\Csf}
\def\Nbb{\Isf{}\kern-1pt{}\Nsf}
\def\Rbb{\Isf{}\kern-1pt{}\Rsf}
\def\Zbb{\Zsf{}\kern-4pt\Zsf}

\def\tav{
			{\hbox{\kern0pt
			\rule[0pt]{2pt}{.5pt}{\kern-3.5pt}
			\rule[0pt]{.5pt}{6pt}{\kern-3.6pt}
			\rule[5.8pt]{4pt}{.5pt}{\kern-3.6pt}
			\rule[0pt]{.5pt}{6pt}
			}}}
\def\tavs{
			{\hbox{\kern0pt
			\rule[0pt]{1.4pt}{.4pt}{\kern-3.5pt}
			\rule[0pt]{.4pt}{4.8pt}{\kern-3.5pt}
			\rule[4.7pt]{3pt}{.4pt}{\kern-3.5pt}
			\rule[0pt]{.4pt}{4.8pt}{\kern-1pt}
			}}}
\def\hbar{h\kern-5pt\vrule width 3pt height 5.75pt depth-5.5pt\kern3pt}


\def\Ac{\hbox{\cal A}}
\def\Bc{\hbox{\cal B}}
\def\Cc{\hbox{\cal C}}
\def\Dc{\hbox{\cal D}}
\def\Fc{\hbox{\cal F}}
\def\Hc{\hbox{\cal H}}
\def\Lc{\hbox{\cal L}}
\def\Mc{\hbox{\cal M}}
\def\Nc{\hbox{\cal N}}
\def\Pc{\hbox{\cal P}}
\def\Qc{\hbox{\cal Q}}
\def\Rc{\hbox{\cal R}}
\def\Sc{\hbox{\cal S}}
\def\Tc{\hbox{\cal T}}
\def\Vc{\hbox{\cal V}}
\def\Xc{\hbox{\cal X}}
\def\Zc{\hbox{\cal Z}}


\def\Arm{{\rm A}}
\def\Arm{{\rm A}}
\def\brm{{\rm b}}
\def\Brm{{\rm B}}
\def\crm{{\rm c}}
\def\Crm{{\rm C}}
\def\Drm{{\rm D}}
\def\Erm{{\rm E}}
\def\erm{{\rm e}}
\def\Frm{{\rm F}}
\def\grm{{\rm g}}
\def\Grm{{\rm G}}
\def\Hrm{{\rm H}}
\def\irm{{\rm i}}
\def\Irm{{\rm I}}
\def\krm{{\rm k}}
\def\Krm{{\rm K}}
\def\Lrm{{\rm L}}
\def\mrm{{\rm m}}
\def\Mrm{{\rm M}}
\def\Nrm{{\rm N}}
\def\nrm{{\rm n}}
\def\Orm{{\rm O}}
\def\orm{{\rm o}}
\def\Prm{{\rm P}}
\def\qrm{{\rm q}}
\def\Qrm{{\rm Q}}
\def\rmq{{\rm q}}
\def\Rrm{{\rm R}}
\def\srm{{\rm s}}
\def\Srm{{\rm S}}
\def\Trm{{\rm T}}
\def\trm{{\rm t}}
\def\Urm{{\rm U}}
\def\vrm{{\rm v}}
\def\Vrm{{\rm V}}
\def\Xrm{{\rm X}}
\def\xrm{{\rm x}}
\def\yrm{{\rm y}}
\def\zrm{{\rm z}}
\def\Zrm{{\rm Z}}

\def\bbox{\vrule height5pt width5pt depth0pt}


\def\dag{\dagger}
\def\ddag{\ddagger}
\def\ox{\otimes}
\def\opl{\oplus}
\def\Oplus{\bigoplus}
\def\plusdotx{{\dot +}}
\def\rx{\stackrel{\rightarrow}{\times}}
\def\lx{\stackrel{\leftarrow}{\times}}
\def\rox{\stackrel{\rightarrow}{\ox}}
\def\lox{\stackrel{\leftarrow}{\ox}}
\def\from{\leftarrow}
\def\Vee{\bigvee}
\def\ox{\otimes}
\def\Ox{\bigotimes}
\def\neqv{\not\equiv}
\def\x{\times}
\def\Bar{{\Big |}}
\def\bra{\langle}
\def\Bra{{\Big\bra}}
\def\ket{\rangle}
\def\Ket{{\Big\ket}}
\def\lst{\left/}
\def\rst{\right/}
\def\lr{\left(}
\def\rr{\right)}
\def\lst{\left|}
\def\rst{\right|}
\def\lss{\left\|}
\def\rss{\right\|}
\def\lc{\left\{}
\def\rc{\right\}}
\def\lb{\left[}
\def\lbb{\sqsubset}
\def\rb{\right]}
\def\rbb{\sqsupset}
\def\cl{\centerline}

\def\Diff{\mathop{\rm Diff}\nolimits}
\def\ISL{\mathop{\rm ISL}\nolimits} 
\def\ISO{\mathop{\rm ISO}\nolimits}
\def\POINCARE{\mathop{\hbox{\rm{P{\scriptsize{OINCAR\'E}}}}}\nolimits}
\def\SL{\mathop{\rm SL}\nolimits}
\def\SO{\mathop{\rm SO}\nolimits}
\def\Spin{\mathop{\hbox{\rm Spin}}\nolimits}
\def\SU{\mathop{\rm SU}\nolimits}
\def\tr{\mathop{\rm tr}\nolimits}
\def\vac{\mathop{\rm vac}\nolimits}

\def\oneb{\hbox{\bf \rm 1}}
\def\twob{\hbox{\bf \rm 2}}
\def\threeb{\hbox{\bf \rm 3}}
\def\fourb{\hbox{\bf \rm 4}}

\def\EQ{\begin{equation}}
\def\ENDEQ{\end{equation}}

\title{
Hypercrystalline vacua}
\author{
David Ritz Finkelstein\thanks
{School of Physics, Georgia Institute of Technology,
Atlanta Georgia 30332.
Partially supported by
NSF
grant PHY-9211036.
E-mail: david.finkelstein@physics.gatech.edu}
\and
Heinrich Saller
\thanks{Heisenberg Institute of Theoretical Physics,
Munich, Germany D-80805. E-mail:
 hns@dmumpiwh.bitnet}
\and
Zhong Tang\thanks{School of Physics,
Georgia Institute of Technology,
Atlanta Georgia 30332.
E-mail: zhong.tang@physics.gatech.edu
Supported by the
M. \& H. Ferst Foundation.}
}
\topmargin=-2cm
\maketitle

\begin{abstract}
Let a quantum network be a Fermi-Dirac assembly
of Fermi-Dirac assemblies of ...of quantum points,
interpreted topologically.
The simplest quantum-network vacuum modes with
exact conservation of
relativistic energy-momentum and angular momentum
also support the local non-Abelian groups
of the standard model, torsion and gravity.
The
spacetime points of these models naturally
obey parastatistics.
We construct a left-handed
vacuum network among others.

\end{abstract}

\section {Vacua as condensates}\label{sec:INTRO}

We work at a level deeper than field theory,
where both spacetime and fields resolve into
a topological network of points and links,
subject to strong locality and q (quantum) superposition.
A vacuum network is a mode (vector) of high symmetry,
a q four-dimensional crystal,
or ``hypercrystal''.
Physical continua resolve into
q superpositions
of discrete modes;
classical spacetime coordinates
and fields are parameters of coherent states.
Unlike Newton's crystalline ether, the vacuum
hypercrystal
is Poincar\'e invariant due to q superposition.
The gauge and matter fields are not extrinsic occupants
but intrinsic excitations or defects
of this
q spacetime network,
to which the hypercrystal is transparent or even
superconducting.
Meissner-Higgs-type effects
concentrate certain imperfections
into
two-dimensional sheets,
the flux sheets of gravity and the other gauge forces
of present-day physics.
The standard-model
symmetries are thus keys to the
hypercrystallography of the vacuum.

The initial results are unexpectedly simple.
Any q hypercubic lattice
is Poincar\'e invariant and has cell
groups corresponding to the standard model,
torsion and gravity
and no other non-abelian gauges.
Thus q topological kinematics and dynamics may
turn out to account for all the forces
rather naturally.

Specifically, we assume here that
the (spacetime-matter) network
is a set of ``links'', which
are sets of points.
This is
simply a ``third-quantized''
Fermi-Dirac assembly of
Fermi-Dirac assemblies.
The kets of q spacetime points,
not analyzed further at this stage,
form a linear space $\Pc$ of high dimension
$|\Pc|\to \infty$\,.
We posit nearest neighbor interactions only.

General covariance was not truly general,
but respected diffentiability.
A truly-general covariance
permits arbitrary point-permutations.
We posit a {\it quantum covariance}
under $\SL(\Pc)$, which includes these
as basis permutations.
This
activates the topology
as Einstein's less-general covariance
activated the metric.
Metric variations accounted for gravity;
we require topological ones to account for all the forces.

The ``spacetime code'' and other early forms of
quantum network dynamics (qnd)\footnote
{Finkelstein, D. (1996).
{\it Quantum Relativity\,}. Springer, Heidelberg.
And references cited there.\label{note:QR}}
fixed the
cell structure at the start.
This
violates
q covariance.
The ``cell'' of a manifold is infinitesimal
and is respected by diffeomorphisms
but the qnd cell is finite.
The qnd dynamics
must govern the spacetime cell structure too,
just as one dynamics
governs atoms, molecules and crystals.
Such a multilevel dynamics calls for a multilevel
tensor analysis. We summarize next the one used here.

\section{Metatensors and parastatistics}
Metatensors are tensors over tensors over $\dots$\,.
We use them
where a classical theory uses sets of sets of $\dots$\,.
Assume
Fermi-Dirac statistics, generalizations being clear.
The quantum algebra $\Qb\Vc$ over any vector space $\Vc$
with dual $\Vc^{\dag}$
is the Clifford algebra
over the normed linear space $\Vc\oplus\Vc^{\dag}$
with norm $||v\oplus \w||:=\w(v)$.\footnote{
H. Saller, Quantum algebras I, II.
{\it Il Nuovo Cimento} {\bf 108B}, 603, {\bf 109B}, 255.}
If $\Vc$ is the ket space of a system $S$
then its quantum algebra
consists of the linear operators
of an (F-D) assembly of $S$'s.
By the {\it metaquantum} algebra over $\Vc$ we mean
$\Qc\Vc:=\lim_{L\to\infty}{\Qb}^L \Vc$ \,.
Its elements we call metatensors over $\Vc$\,.
The hypothetical q system with ket space $\Qc\Vc$
we call the universal quantum over $\Vc$\,.
It
is self-referential in much the way that set theory is:
$\Qc\Vc$ includes all (sufficiently finite) operators on itself.

Tensors have grade $g$. Metatensors
have level $L$,
counting  unitizations up from $\Vc\oplus\Vc^{\dag}$\,
and a grade $g^L$ at each level.
They also have constituents at any depth $D \le L$
counting down from $L$ and a grade $g_D$ at each depth.

Here we take the qnd network to have as its quantum algebra
the universal quantum algebra
$\Qc:=\Qc\{0\}$ and
basis $\N_Q=|Q\ket$\,.
Then qnd kets are first-grade elements of $\Qc$.
Link kets in turn are depth-one grade-one factors of qnd kets.
Spacetime point kets are depth-two grade-one
factors of qnd kets.
For any depth $D\in\Nbb$\,, let $c^D_Q$
be the creation operator on $\Qc$
creating factors $|Q\ket$ at depth
$D$ with
dual annihilator $\partial^Q_D$\/.
We identify quantum covariance with invariance under
the group generated by all $L^Q{}_{P} c^1_Q \partial_1^P$
with $\tr L=0$\,.

The action operator $\tilde S$ acts on $\Qc$\,,
hence ${\tilde S}\in\Qc(\Pc)$.
Qnd kets $\Psi$
(e.g., $|\vac\ket$) must obey the
differential subsidiary conditions
$[\partial^1_Q \tilde  S|\Psi=0$\,.

Typical action terms:
\begin{itemize}
\item{$c_A \partial^A =$ network number, modulus}
\item{$c^1_Q \partial_1^Q\sim$ link number}
\item{$c2_Q \partial_1^Q\sim$ point number}
\item{$c^2_Qc^{1}[^Q{}{_P}]\partial_2^{P}\, \dots$  $\sim$
first-neighbor point-interaction}
\end{itemize}
with
composite link index $[^Q{}_P]$\,.

Points obey parastatistics.
This is an  unintended immediate
consequence of qnd, not a separate physical hypothesis.
A product extensor in $\Qb^2(\Pc)$
must change sign when two of its (first-grade!) factors
are interchanged,
but need not change sign
when two subfactors of its factors are interchanged.
This simple quantum fact has
a simple classical analogue:
Any set (for example, $\{\{1,2\},\{3,4\}\}$) is
invariant under an interchange
of two elements (say, $\{1,2\}$ and $\{3,4\}$\,),
but not under the interchange of {\em their} elements
{say, 1 and 3).

\section{Metrical vacua}

We turn now to empirical hypercrystallography of the vacuum.
Assume four commuting
coordinate operators $n^{\m}$ on a point-ket
space $\Pc$\,, each with spectrum $\Zbb$
(not  $\Nbb$
as in earlier work$^{\ref{note:QR}}$),
defining a hypercubical array of eigenkets
\EQ
|n^1, n^2, n^3, n^4 \ket=|n\ket\in \Pc\,,\quad
\bra n^1, n^2, n^3, n^4 |=\bra n|\in \Pc^{\dag}\,.
\ENDEQ
These are not fundamental but
descend from
the higher level of dynamics
by spontaneous $\SL(\Pc)$-breaking,
and give rise in turn to the still-less-fundamental
points and coordinates
of Minkowski spacetime as follows.
We identify the spacetime translation
generators $\partial_\m$ with
down-shift operators (not antihermitian!)
and the point coordinate operators
with upshift operators,
\begin{eqnarray}\label{eq:pam}
\partial_{\m}:=\sum_n &|n-1_{\m}\ket\vee\bra n|&,\cr
x^{\m}:=\sum_n &|n+1_{m}\ket\vee\bra n|&\,(n+1_{\m})\,,
\end{eqnarray}
scaled to save the commutation relations
of differential geometry,
$
[\partial_{\m}, x^{\l}]=\d^{\l}_{\m}\,.
$
It is then easy to see that q covariance
includes general covariance,
$\SL(\Pc) \supset$ $ \Diff(\Rbb^4)$ $ \supset \ISO(1,3)$\,.
Namely, the representative $\d\L:\Pc \to \Pc$ of the
infinitesimal diffeomorphism
$\d x=(\d x(x)): \Rbb^4\to \Rbb^4$ is
$\d \L=-\d x\cdot\partial\,.$
To represent $\POINCARE=\ISO(1,3)$, one sets
$\d x^\m=\a^{\m}+\w^\m{}_{\n}x^{\n}$\,
with infinitesimal parameters $\a$, $\w$
with $\w_{\m\n}=-\w_{\n\m}$\,.

We propose some
suitably symmetric patterns of topological linkages
among these point-kets for vacuum  hypercrystals.
One early$^{\ref{note:QR}}$ model
was a four-dipole metatensor
with off-diagonal long-range order,
$|\vac\ket=\i\partial_1 \vee \dots \vee \i\partial_4$\,.
Like a hypervolume element, this is clearly
invariant under the above non-unitary
Poincar\'e subgroup of $\Diff$\,.
Its unit cell has $4!=2\x 3\x 4$ bonus discrete symmetries
that we tentatively identified with generators of
quark hypercharge, isospin, color and spin:
mutually-commuting coherent-state
$\Urm_1, \SU_2$, $\SU_3$,
and $\Spin_4$ groups.$^{\ref{note:QR}}$

This vacuum, however, is {\em too} symmetric.
Like a hypervolume element, it is invariant
under the inhomogeneous special linear group
$\ISL(4)\supset\ISO(1,3)$\,.
It defines the volume element but not the metric.
We construct three metrical vacua here.

The ``Dalembertian vacuum'' is the metatensor \footnote
{This improves on an earlier ``quadrupole vacuum'' of
D. Finkelstein, H. Saller, and Z. Tang,
Beneath gauge, {\it Nuclear Physics}\,,
to appear in the Trautman issue.}
\EQ
|\vac\ket\sim \i g^{\m\n}\i[\partial_{\m}\circ\partial_{\n}]\,.
\label{eq:DAV}
\ENDEQ
The 4! discrete symmetries respect this for
$g^{\m\n}=1-\d^{\m\n}$ (null symmetric form).
The ``Dirac vacuum'' is a square root of the Dalembertian vacuum:
\EQ
|\vac\ket\sim\i[\gamma^\m \vee\i\partial_\m]\,,
\label{eq:DIV}
\ENDEQ
The ``left-handed vacuum'' is a restriction of the Dirac vacuum:
\EQ |\vac\ket\sim\i[\sigma^\m \vee \i\partial_\m]
\label{eq:LHV}
\ENDEQ
(\ref{eq:DAV}) uses the $\partial$ given by (\ref{eq:pam}).
(\ref{eq:DAV}) uses the operator product $\circ$
because the
Grassmann product would give 0.
(\ref{eq:DIV}) uses a K\"ahler representation
of the Dirac $\gamma$ operators by network
metatensors defined
elsewhere$^{\ref{note:QR}}$.
(\ref{eq:LHV}) uses a $\gamma^5$
superselection
rule, Hestenes' identification $\gamma^5=i$,
and $\sigma^\m\sim\gamma^\m$\,.
All three vacua have the local standard-model unitary
symmetries and global proper Poincar\'e
symmetries; only the left-handed vacuum
lacks parity.

These quasi-empirical hypercrystalline vacua
conserve angular momentum and
energy-momentum but not shear and dilation.
Their spacetime points automatically exhibit parastatistics.
Their cells have
algebras isomorphic to
the local algebras of gravity, torsion and the standard
model. One violates parity.
The chances of one of these guessed vacua
being right are nevertheless
slim.
They simply show that if we take quantum theory seriously,
we can make quite simple locally finite
hypercrystalline vacua that have the same
symmetries as the singular classical continua
used in the usual quantum field theory,
which arbitrarily restricts quantum theory to one level.
We must successively correct the vacuum structure
and the action in turn to
converge to a working theory.

\end{document}